\documentclass[a4paper,11pt]{article}
\pdfoutput=1
\usepackage{jinstpub}
\usepackage{setspace}
\usepackage[export]{adjustbox}
\usepackage{lineno}
\usepackage[normalem]{ulem}
\usepackage{color}
\definecolor{blue}{rgb}{0,0,1}
\definecolor{red}{rgb}{1,0,0}

\begin{document}

%% Title page.
\title{\boldmath The Forward TPC system of the NA61/SHINE experiment at CERN: a
  tandem TPC concept}

\author[a]{Brant Rumberger,}
\author[c,\dagger]{Antoni Aduszkiewicz,}
\author[a]{Jan Boissevain,}
\author[c]{Magdalena Kuich,}
\author[b]{Andr\'as L\'aszl\'o,}
\author[a]{Yoshikazu Nagai,}
\author[b,*]{L\'aszl\'o Ol\'ah,}
\author[c]{Piotr Podlaski,}
\author[b]{Dezs\H o Varga,}
\author[d]{Martin Wensveen,}
\author[a]{Eric D.\ Zimmerman}

\affiliation[a]{University of Colorado Boulder, Boulder, USA}
\affiliation[b]{Wigner Research Centre for Physics, Budapest, Hungary}
\affiliation[c]{Faculty of Physics, University of Warsaw, Poland}
\affiliation[d]{CERN, Geneva, Switzerland}
\affiliation[*]{{\rm Current address:} University of Tokyo, Tokyo, Japan}
\affiliation[\dagger]{{\rm Current address:} University of Houston, Houston, USA}

\emailAdd{
  brant.rumberger@colorado.edu,
  antoni.aduszkiewicz@cern.ch,
  jgboissevain@comcast.net,
  magdalena.kuich@cern.ch,
  laszlo.andras@wigner.hu,
  yoshikazu.nagai@colorado.edu,
  olah.laszlo@wigner.hu,
  piotr.podlaski@cern.ch,
  varga.dezso@wigner.hu,
  martin.wensveen@cern.ch,
  edz@colorado.edu
}

\abstract{ This paper presents the Forward Time Projection Chamber
  (FTPC) system of the NA61/SHINE experiment at the CERN SPS
  accelerator.  This TPC system applies a novel tandem-TPC design to
  reduce the background originating from particle tracks
  not synchronous with the event trigger.  The FTPC system is composed
  of three chambers with alternating drift field directions.  The
  chambers were installed directly along the beamline region of the
  NA61/SHINE detector in a medium- to high-intensity
  ($10-100\,\mathrm{kHz}$) hadron or ion beam. The tandem
  TPC system has proved to be capable of rejecting out-of-time
  background tracks not associated with a primary interaction.
  In addition,
  the system performs tracking and inclusive
  $\mathrm{d}E/\mathrm{d}x$ particle identification for particles
  at and near the beam momentum. This
  shows that a tandem-TPC-based chamber design may be used also in
  other experimental applications with a demand for low material
  budget, tracking capability, and the need for $\mathrm{d}E/\mathrm{d}x$
  particle identification, all while in the presence of a relatively high
  particle flux.  }

% Only keywords from JINST's keywords list please

\arxivnumber{2004.11358}

\maketitle
\flushbottom

\section{Introduction}
\label{introduction}

NA61, also known as the SPS Heavy Ion and Neutrino Experiment (SHINE) 
\cite{abgrall2014}, is a large-acceptance hadron spectrometer experiment that recieves
beam from the CERN Super Proton Synchrotron (SPS). The original design of the NA61 tracking
system, mainly optimized for 
heavy-ion physics, lacked full phase space 
coverage in the forward region. 
Complete coverage of this part of phase space is of particular importance to the neutrino 
area of the NA61 physics program, which provides hadron production 
measurements of neutrino ancestors to long-baseline neutrino oscillation 
experiments. Covering this missing region of phase space with a tracking 
detector capable of performing particle identification is crucial for 
making accurate neutrino flux predictions for such experiments.

\begin{figure}[!h]
  \begin{center}
    \includegraphics[height=4.4cm]{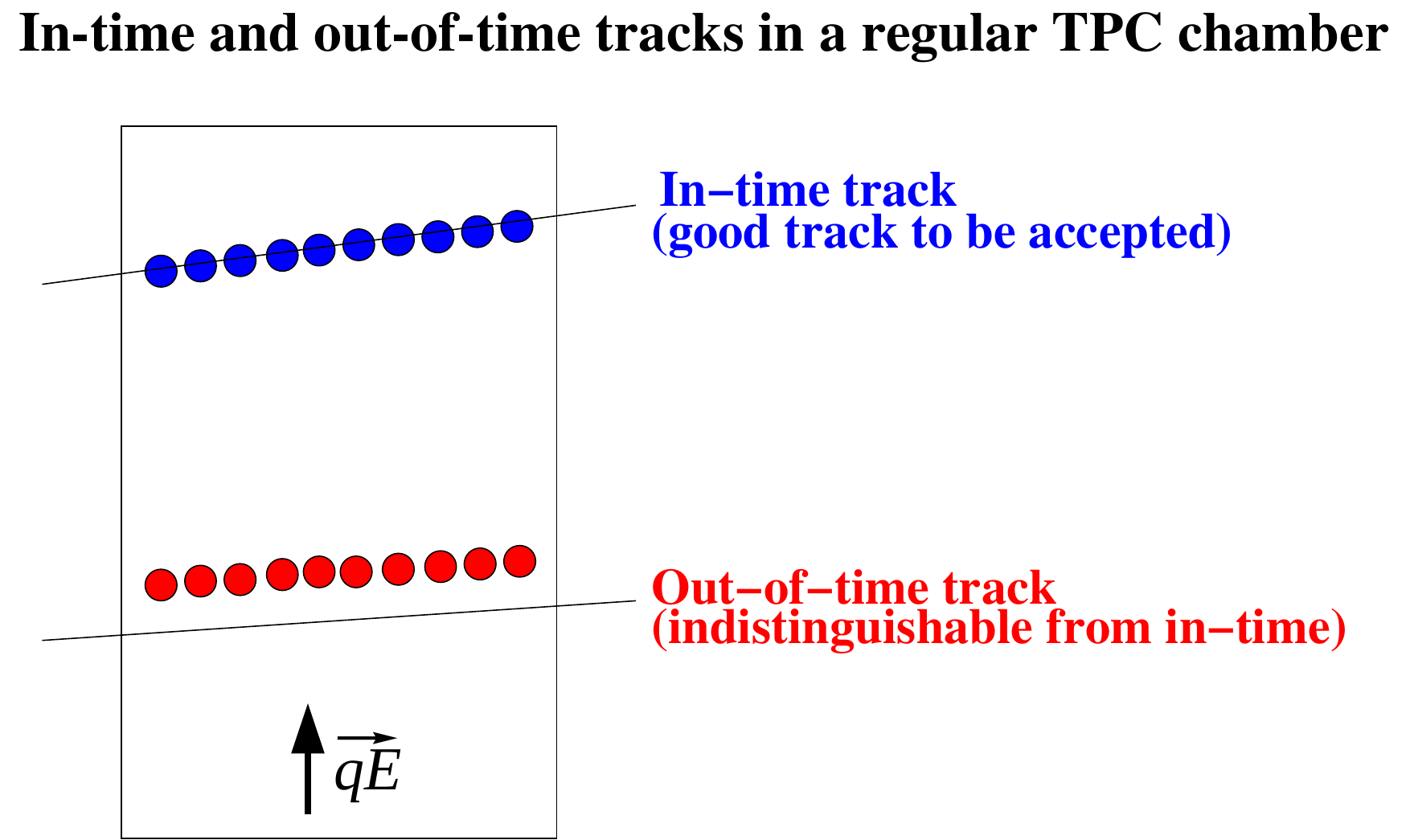}$\qquad$%
    \includegraphics[height=4.4cm]{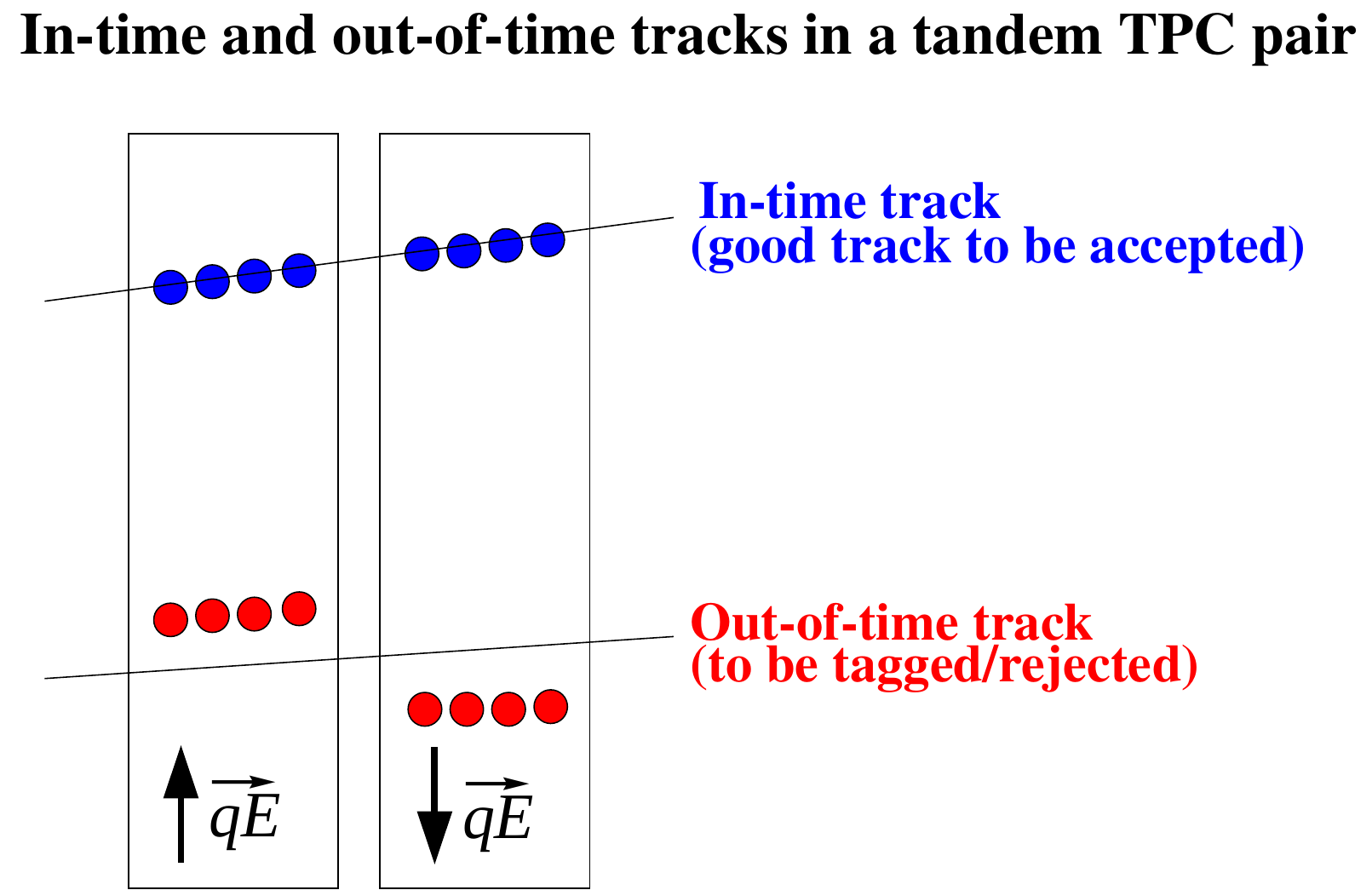}
  \end{center}
  \caption{(Color online) Illustration of the tandem-TPC concept. 
    Left panel: in a regular TPC, in-time tracks and the out-of-time particle tracks are 
    indistinguishable.
    Right panel: in a tandem-TPC pair the drift directions ($q\vec{E}$) are not the same 
    in the two drift volumes, and therefore the segments of an out-of-time 
    particle track already start to drift apart by the time of the arrival 
    of the interaction trigger. 
    Thus, by their discontinuity, the out-of-time particle tracks become 
    distinguishable from true in-time tracks.
  }
  \label{figTandemConcept}
\end{figure}

In this paper, we describe a time projection chamber (TPC) detector 
system used for instrumenting this missing region of phase space of the 
NA61 experiment. Since this missing region encompasses the beam, and since
not all beam particles result in a physicswise interesting event, the separation of the 
true in-time tracks from the out-of-time particle tracks (not synchronous to 
the interaction trigger) becomes essential.\footnote{We call a particle 
\emph{in-time} if it originates from the 
interaction on which the event was triggered. A particle is called \emph{out-of-time} 
if it originates from a previous collision, not synchronous to the interaction 
trigger, and therefore ideally should be rejected or tagged. 
Since TPC chambers in general are intrinsically relatively slow devices, 
with a typical total drift time of the order of a few $10$ $\mu$secs, TPC chambers 
close to the beam region are generally contaminated with out-of-time tracks 
if the beam intensity is not very low.}
In order to address the recognition of out-of-time tracks, a TPC concept was proposed 
consisting of separate drift volumes with alternating drift field directions. 
We call this the \emph{tandem-TPC} concept. The idea is illustrated in 
Figure~\ref{figTandemConcept}. Specifically, our design incorporates three 
separate consecutive TPCs (plus a fourth that already existed) with alternating drift fields. 
The chambers are collectively referred to as the Forward Time Projection 
Chambers (FTPCs), as they cover the most forward region of the phase space 
of the NA61 experiment.

Due to a relatively long total drift time 
(${\sim}50\,\mu\mathrm{sec}$) of the TPC chambers, 
and a beam intensity of $10\,{-}\,100\,\mathrm{kHz}$, 
the presence of out-of-time particles is rather 
common in a typical NA61 event. These out-of-time particles can be of 
any species, as the SPS beam used is typically secondary in nature. Out-of-time particles
are not subject to the NA61 particle identification 
criteria during triggering, and thus can introduce bias in analyses.

The NA61 FTPCs satisfy additional design constraints imposed by the 
multi-purpose nature of the experiment. In order to preserve the data taking 
environment for the NA61 heavy-ion program, the FTPCs were constructed using 
a novel low material budget technique, using etched copper-coated 
polyimide film  
for the field cages. The film is coated with copper on one side only, resulting in only
$0.3\,\%$ radiation lengths of material. 
In order to maintain a low oxygen contamination within the active volume, 
an additional thin Mylar foil wall envelopes the field cage from the outside. 
The two foils provide a buffer volume, which is flushed by the exhaust gas 
in order to reduce the diffusion from the air into the active volume. 
This design minimized the gap between these two walls, 
in order to use the available detector space economically.

Using TPCs with opposite drift direction has been successfully applied earlier 
for small TPC systems in high rate environments. The Sextant 
\cite{legou2007} was a high precision tracking component, proposed to be 
part of the KABES upgrade of the CERN NA48 experiment, using a Micro-mesh 
readout plane. The ``Twin-configuration GEM TPC'' \cite{garcia2018}, 
with similar geometry as Sextant but using GEM on the readout plane, was 
designed for beam particle tracking from protons up to uranium ions. The 
present paper demonstrates that the ``tandem-TPC'' concept can be used not 
only for large TPCs, but also in a more general setting, having TPCs of 
different sizes and geometries, and with different drift velocities. 
The specific choice presented in this paper, as discussed above, was 
guided by the general environment of the NA61 detector.

\section{The NA61/SHINE experimental facility}
\label{na61ExperimentalFacility}

The NA61/SHINE detector is a multi-purpose fixed-target hadron spectrometer 
\cite{abgrall2014} at the CERN SPS accelerator. 
Large parts of its main tracking devices were inherited 
from a previous experiment called NA49 \cite{afanasiev1999}. Its main 
physics goals are to search for the critical point of
strongly-interacting 
matter, to study the onset of deconfinement in quantum chromodynamics,
and to measure identified particle production 
spectra in hadron-nucleus collisions as reference data for 
long-baseline neutrino experiments and large area cosmic ray 
observatories. 

\begin{figure}[!h]
  \begin{center}
    \includegraphics[width=14cm]{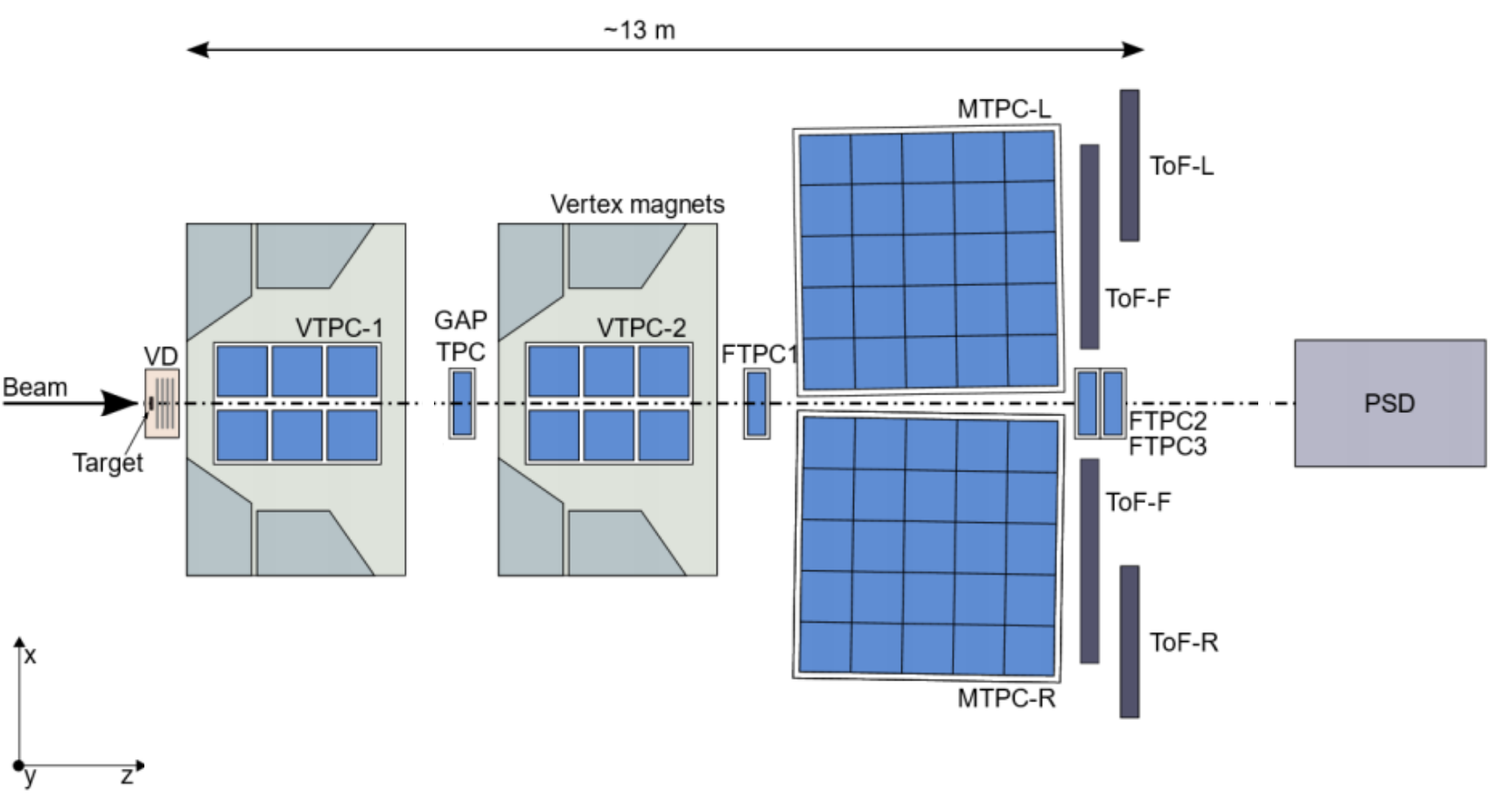}
  \end{center}
  \caption{(Color online) The NA61/SHINE
    (SPS Heavy Ion and Neutrino Experiment) detector configuration,
    with the Forward Time Projection Chambers (FTPCs). Prior to 2017 the GapTPC was the only
    detector capable of measuring particles in the beamline
    region. With only seven measurement planes, the GapTPC
    could not perform accurate $\mathrm{d}E/\mathrm{d}x$ estimation on its own, nor
    could it alone determine the momentum of very forward particles.
    Additional detectors displayed:
    Vertex Detector (VD), Vertex TPC 1 \& 2 (VTPC-1 \& VTPC-2), Main TPC Left \& Right
    (MTPC-L \& MTPC-R), Time-of-Flight wall Left \& Right \& Forward (TOF-L \& TOF-R \& TOF-F),
    Projectile Spectator Detector (PSD).
  }
  \label{figDetectorSetup}
\end{figure}

The outline of the NA61/SHINE experiment is presented in Figure~\ref{figDetectorSetup}. 
Two superconducting bending magnets (Vertex I and II) are responsible for 
particle deflection for momentum determination, with a total maximum bending 
power of ${\sim}9\,\mathrm{Tm}$ (up to $1.5\,\mathrm{Tesla}$ in Vertex I and
$1.1\,\mathrm{Tesla}$ in Vertex II). 
A target holder with target moving capability sits just upstream of the first Vertex 
TPC. Thin targets can be placed inside a silicon vertex detector (VD) upstream of Vertex I
for precise vertex determination.
NA61/SHINE also has the ability to measure interactions in extended 
replica targets for long-baseline neutrino experiments. The tracking devices for spectrometry 
are composed of eight large volume TPCs (total ${\sim}40\,\mathrm{m}^{3}$ and 
${\sim}1\,\mathrm{m}$ drift length), 
capable of performing both tracking and $\mathrm{d}E/\mathrm{d}x$ measurements. 
Three Time-of-Flight walls (ToF-L, ToF-R, ToF-F) 
complete the particle identification (PID) phase space coverage and enable two-dimensional
separation of particle species.
A calorimeter is placed at the end of the beamline, called the Projectile 
Spectator Detector (PSD), which helps to estimate centrality in 
heavy-ion collisions.
Upstream of the target position, a set of scintillator and Cherenkov detectors serves as beam and 
beam PID trigger (not shown on the figure). Between VTPC-1 and GapTPC on the 
beamline, a small plastic scintillator of $1\,\mathrm{cm}$ diameter serves as 
an interaction trigger in most collision types (not shown on the figure).

The pre-existing NA61 TPCs (VTPC-1, VTPC-2, MTPC-L, MTPC-R, GapTPC) 
were designed primarily to meet the experimental needs of the heavy-ion 
program of the NA49 experiment. The Vertex TPCs (VTPC-1 and VTPC-2) sit in the
vertex magnets and each provide $72$ position measurement planes in order to determine
track momentum. They are not instrumented along the beamline due to the high
concentration of charged ion fragments and delta electrons in this region, 
where high levels of ionization would make wire plane instrumentation difficult. 
The Main TPCs (MTPC-L and MTPC-R) provide $90$ additional measurement planes
for tracks, enabling accurate particle identification 
via $\mathrm{d}E/\mathrm{d}x$ and increasing the tracking lever arm. 
The removable GapTPC was added to in order to improve the momentum and vertex 
resolution of MTPC-only tracks, and also in order to have some charged particle 
detection capability in the beamline region, for non-heavy-ion measurements. 
The GapTPC contains only seven detection planes, and on its own cannot 
accurately perform momentum and $\mathrm{d}E/\mathrm{d}x$ measurements for 
particles near the beam momentum.

The shortcomings of the original tracking and PID capabilities near the  
beamline of the spectrometer became a bottleneck for the 
NA61/SHINE physics program related to long-baseline neutrino experiments. 
In order to close the pertinent missing tracking and PID acceptance gap, 
the FTPCs (FTPC-1, FTPC-2 and FTPC-3) were constructed in 2017. 
FTPC-1 is located downstream of the second Vertex magnet, and FTPCs 2 and 3 
are located downstream of the MTPCs. Each FTPC contains $12$ measurement 
planes, giving $36$ total tracking and $\mathrm{d}E/\mathrm{d}x$ points. 
%Since the FTPC chambers reside directly in the beamline, special 
%measures had to be taken for background rejection: each adjacent FTPC drift 
%volume has opposite drift direction, which helps to identify particle 
%traces not associated to a recorded collision (tandem-TPC concept for 
%out-of-time particle trace rejection).

\section{The FTPC design concepts}
\label{design}

When selecting a detector design and technology for instrumenting the 
forward region of NA61, several performance goals were considered:
\begin{itemize}
\item low material budget in order not to disturb performance of MTPC and PSD,
\item tracking capability for covering the missing acceptance,
\item PID capability, meaning at least inclusive proton-pion separation,
\item powerful out-of-time track rejection capability,
\item compatibility with the existing NA61 infrastructure,
\item cost-effectiveness.
\end{itemize}

The low material budget assures that the new forward detectors do not 
detract from other physics programs: the first chamber is situated upstream of 
the MTPCs, which provide crucial $\mathrm{d}E/\mathrm{d}x$ samples along 
particle trajectories. The MTPC track samples need to be biased as little as
possible in order to preserve the existing 
$\mathrm{d}E/\mathrm{d}x$ resolution of NA61. 
Furthermore, downstream of the FTPC-2 and 3 stations the PSD calorimeter 
resides for detection of heavy-ion fragments. Thus, excessive
material had to be avoided upstream of the PSD calorimeter as well, in 
order not to initiate premature electromagnetic or hadronic showers.

The inclusive proton-pion separation capability proved to be necessary 
in order to make the planned production measurements useful for the 
long-baseline neutrino experiments. The design goal of the forward detectors was 
to achieve statistical separation of protons and pions at $100\,\mathrm{GeV/}c$ 
forward momentum. In the NA61 configuration, this momentum is within the 
relativistic rise region of the Bethe--Bloch curve for Ar/CO$_2$ gas mixture, 
which is the working gas for the existing NA61 TPCs 
(for VTPCs 90:10 Ar/CO$_2$ is used, whereas for all the other TPCs 95:5 mixture is used). 
This fact was one of the 
main motivations for choosing the TPC concept as a detector technology, as
a TPC is capable of implementing inclusive proton-pion separation via 
$\mathrm{d}E/\mathrm{d}x$ to a sufficient quality at a relatively low cost.

The issue of out-of-time track rejection was foreseen to be crucial due 
to the position of the detectors in the medium-intensity beamline.
A typical $4\,\mathrm{sec}$ long 
spill contains $10^{5} - 10^{6}$ beam particles, while the total TPC drift 
time in NA61 is $51.2\,\mu\mathrm{s}$ over the ${\sim}1\,\mathrm{m}$ 
of drift distance. The beam particles in a typical 
spill are not necessarily uniformly spaced in time --- the spill structure 
can vary significantly due to changes in the accelerator chain. Non-uniformity 
in the spill time structure results in significant intensity fluctuations, 
causing a large fraction of collected events to contain more than $10$ 
out-of-time beam particles. Moreover, upstream collisions in the 
beamline also contribute to the yield of out-of-time particles close to beam 
momentum.

Detector compatibility is an important point for selecting a detector 
technology. The forward region has limited physical space due to the presence of 
the existing NA61 TPCs. Existing infrastructure such as the Ar/CO$_2$ gas supply 
were also be taken into consideration. Data Acquisition System (DAQ) and 
software compatibility also influenced the design decisions.

The tandem-TPC concept was selected in order to satisfy these
requirements. Advantages of the resulting system, with the 
specific choice of the solutions, are the following:
\begin{itemize}
\item Low material budget field cage constructed from copper-coated 
  polyimide foil.
\item 36 FTPC + 7 GapTPC $\mathrm{d}E/\mathrm{d}x$ points for effective
  proton-pion separation at $100\,\mathrm{GeV/c}$.
\item Tandem field cage design for out-of-time track rejection.
\item Backward compatibility with existing NA61 front-end electronics and gas system.
\item Relatively low-cost detector technology and construction.
\end{itemize}

Two solutions ensured a low contribution to the overall detector material
budget. First, the field cage was constructed using a $75\,\mu\mathrm{m}$ thick 
polyimide
film bonded to an $18\,\mu\mathrm{m}$ layer of copper foil. The foil was etched into 
$4.5\,\mathrm{mm}$ parallel strips spaced $0.5\,\mathrm{mm}$ apart serving as the equipotential surfaces
of the field cage, which are joined by low-profile $1\,\mathrm{MOhm}$ resistors. The
field cage was surrounded by a layer of $50\, \mu\mathrm{m}$ thick Mylar foil, providing
the outer seal for the gas volume. Overall, the field cage and Mylar windows
comprise $0.34\,\%$ radiation lengths. The active Ar/CO$_2$ (95:5) gas volume comprises
$0.41\,\%$ radiation lengths.  Second, the upstream FTPC was designed to be
removable during heavy ion data taking periods. The FTPC1 support structure
includes a sliding rail system and easily detachable chamber infrastructure,
allowing the entire chamber to be moved out of NA61 active region, when not used. 
The downstream FTPCs were not designed to be removable, since they are installed
downstream of the MTPCs. When out of use, they can be filled with helium in order to reduce
material shadowing of the PSD.

In order to achieve statistical separation of protons and pions at $100\,\mathrm{GeV/c}$,
multiple $\mathrm{d}E/\mathrm{d}x$ samples must be collected along particle
tracks. NA61 TPC front-end electronics allow for charge measurement at each
measured position. The FTPC system segments the tracks into $12$ point and
charge measurements per chamber, giving $36$ total measurements over $144\,\mathrm{cm}$
of instrumented track length. 
The existing beamline TPC, the GapTPC, provides $7$ additional measurements, totaling to $43$
$\mathrm{d}E/\mathrm{d}x$ points. Using the relation
$\sigma_{\mathrm{d}E/\mathrm{d}x} \sim \frac{1}{\sqrt{N}}$ where $N$ is the
number of measurement points \cite{afanasiev1999}, we have determined that the
instrumentation will be sufficient to statistically separate protons from pions
above $10\,\mathrm{GeV/c}$, well above the Bethe--Bloch crossing for protons and pions in
the used detector gas composition.

Oxygen contamination of the active gas reduces $\mathrm{d}E/\mathrm{d}x$ 
resolution. The contamination in the FTPCs is mitigated by including a 
buffer volume of detector gas around the outside of the sensitive
volume. The field cage forms a gas tight seal with the supply gas, which 
is forced through a hole pattern in the cathode. The flow is then redirected 
near the wire plane up through the gas-tight envelope formed by the backside 
of the field cage and the Mylar window. This design is illustrated in
the right panel of Figure~\ref{figTandemFieldCage}. 
Gas contamination due to diffusion through the Mylar is thus relegated to the exhaust gas, and 
the contamination penetration to the sensitive volume is reduced. 
For larger volume TPCs, many experiments \cite{abgrall2014,afanasiev1999} apply 
an additional gas circuit for constantly flushing the gas envelope around the 
chamber, usually with dry Nitrogen. 
The relatively modest volume of the FTPC chambers allowed us to avoid such a 
complication, by reusing the exhaust gas for filling the buffer volume. 
That simplification in the design also made it possible to make the buffer 
volume rather thin, of the order of $1\,\mathrm{cm}$, allowing for more 
economic usage of the available detector space.

The tandem design was achieved by orienting the field cage drift directions of
the most upstream and most downstream FTPCs (FTPCs 1 \& 3, respectively) in the
downward direction, while orienting the drift direction of FTPC-2 upward. 
The readout wire planes for these two chambers are on the
bottom of the chambers, while for the rest of NA61 and for the center FTPC
(FTPC2) the readout wire planes are on top of the chambers. This creates a tandem
TPC system between four TPCs: GapTPC \& FTPC2 drift directions are oriented in the
${+}y$ direction, while FTPC1 \& FTPC3 drift directions are oriented in the ${-}y$
direction.

The pre-existing NA61 TPC design is mainly an updated version of TPCs inherited
from NA49. Several design aspects were selected in order for the new FTPCs to be
compatible with existing detector infrastructure. The wire plane readout is
compatible with the NA61 front-end electronics (FEE), allowing for the FTPCs to be easily integrated
with the existing detector DAQ system. The gas volume accepts an identical gas
mixture to the one used in the MTPCs (Ar/CO$_2$, 95:5), allowing for smooth
integration into the existing infrastructure. Additional drift velocity monitors
similar to ones for existing TPCs were constructed and installed in the
experiment \cite{kuich2019}.

The FTPCs were constructed at two different facilities in two logical parts: 
the field cage and gas volume, and the amplification and readout wire plane. 
The design concept of the field cage and gas volume was largely motivated by 
an earlier small volume TPC chamber, called the LMPD \cite{marton2014}. 
The actual real scale design and prototyping was done at the 
University of Colorado Boulder in the USA, and the 
assembly was done at CERN in Switzerland/France, due to the problems 
of transportation of a fragile pre-assembled field cage. 
The readout and amplification wire plane was constructed at the Wigner Research Centre 
for Physics, Budapest, Hungary, and was subsequently shipped to CERN, since 
a pre-assembled wire plane proved to be strong enough for transportation. 
The above strategy necessitated a modular structure, such that the above two 
main components could be constructed and tested independently, and eventually 
joined together at CERN.

\section{Engineering solutions}
\label{engineering}

\subsection{Field Cage}
\label{fieldcage}

\begin{figure}[!tbp]
  \begin{center}
    \includegraphics[width=\linewidth]{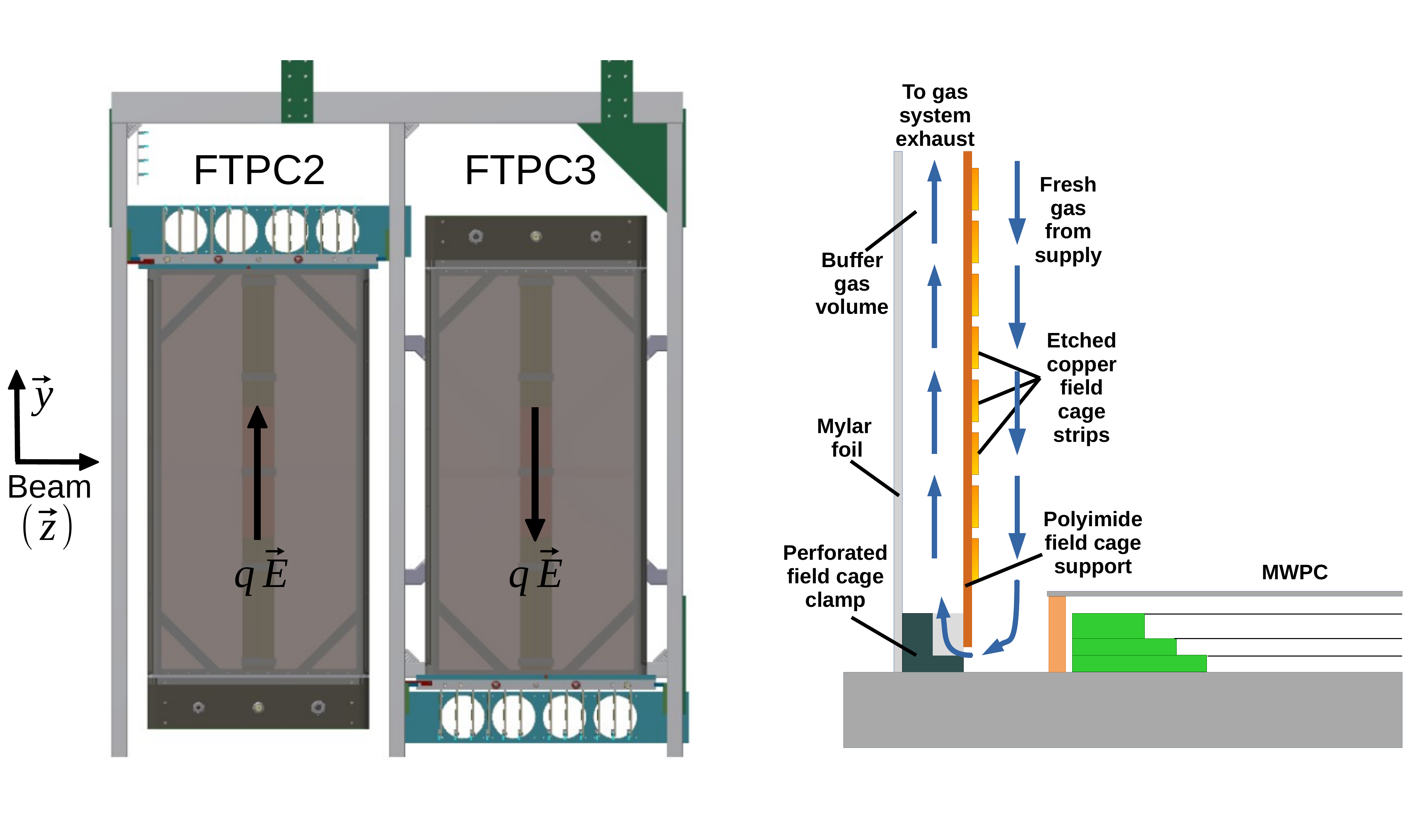}
  \end {center}
  \caption{(Color online) Left: Mechanical realization of the tandem-TPC concept 
    in the FTPC2-FTPC3 tandem pair chambers. The actual drift field directions 
    $q\vec{E}$ along with the NA61 coordinate axis convention is also 
    depicted in the figure. Right: Detail of the buffer volume and field cage
    near the readout plane. Gas flow is shown in blue.
  }
  \label{figTandemFieldCage}
\end{figure}

The main TPCs are shadowed by not only the active areas of the FTPC1
field cage but also the support posts for the field cage. These must
therefore be made with minimal material, unlike the rigid posts that
support tensioned field cage strips in the older chambers.  The
single-sheet copper plated polyimide field cage is therefore supported
under minimum tension by a four-windowed
Noryl\textsuperscript{\textregistered} frame manufactured at the
University of Colorado Boulder. This support structure consists of
four $1\,\mathrm{m}$ tall C-shaped support posts. The polyimide side
of the field cage is adhered to the inside of the support posts, and
the outer Mylar window is adhered to the outside of the posts. The
single foil field cage solution avoids issues related to fine
positioning of field cage strips, as the copper strips are situated
directly on the sheet of polyimide.

A clamping retention mechanism also constructed from
Noryl\textsuperscript{\textregistered} keeps the field cage in place along the
tops and bottoms of the support windows. The top edge of the field cage is
clamped to the cathode retention box, and the bottom edge is clamped to the
aluminium flange used for attachment to the readout wire plane. This system maintains
proper field cage positioning and tension over the windows, which cover an area
of $101\times 71\,\mathrm{cm}^2$ on the longest sides.

The Noryl\textsuperscript{\textregistered} cathode retention box electrically
isolates the $20\,\mathrm{kV}$ cathode from the outside of the chamber and the gas supply
lines. It also houses the gas distribution and exhaust systems. Fresh chamber
gas flows into a PerFluoroAlkoxy (PFA) inlet tube, which in turn forces the gas
through a ``shower head'' hole pattern in the cathode. The use of PFA tubing
as opposed to metallic tubing ensures that the cathode is well-isolated from
any potential grounding routes.
This gas then flows
through the active chamber volume before being forced into the buffer volume
between the outside of the field cage and the outer Mylar window. This gas is
forced back into the cathode retention box and out through an exhaust line.

The fully constructed field cage unit can be sealed with an aluminium sheet
in place of the wire frame assembly, via
insertion of an O-ring in the wire plane attachment flange. This allowed for
independent construction of the field cages, along with gas tightness and
high voltage testing.

The rendering of the mechanical design of the FTPC2-FTPC3 tandem pair is 
presented in the left panel of Figure~\ref{figTandemFieldCage}.

\subsection{Amplification and readout wire plane}
\label{wireplane}

As usual in a TPC setting, the ionization electrons which form along the particle 
track are guided by the homogeneous drift field towards the readout plane, 
which resides at the end of the drift volume.
The technology for the amplification and readout plane was chosen 
to be based on the multi-wire proportional chamber (MWPC) concept. This 
choice was made for the sake of simplicity of design, construction 
and integration with existing NA61 components, 
such as the Data Acquisition (DAQ) \cite{laszlo2015} and the 
offline software (Shine) \cite{sipos2012}. 

\begin{figure}[!h]
  \begin{center}
    \includegraphics[width=0.8\linewidth]{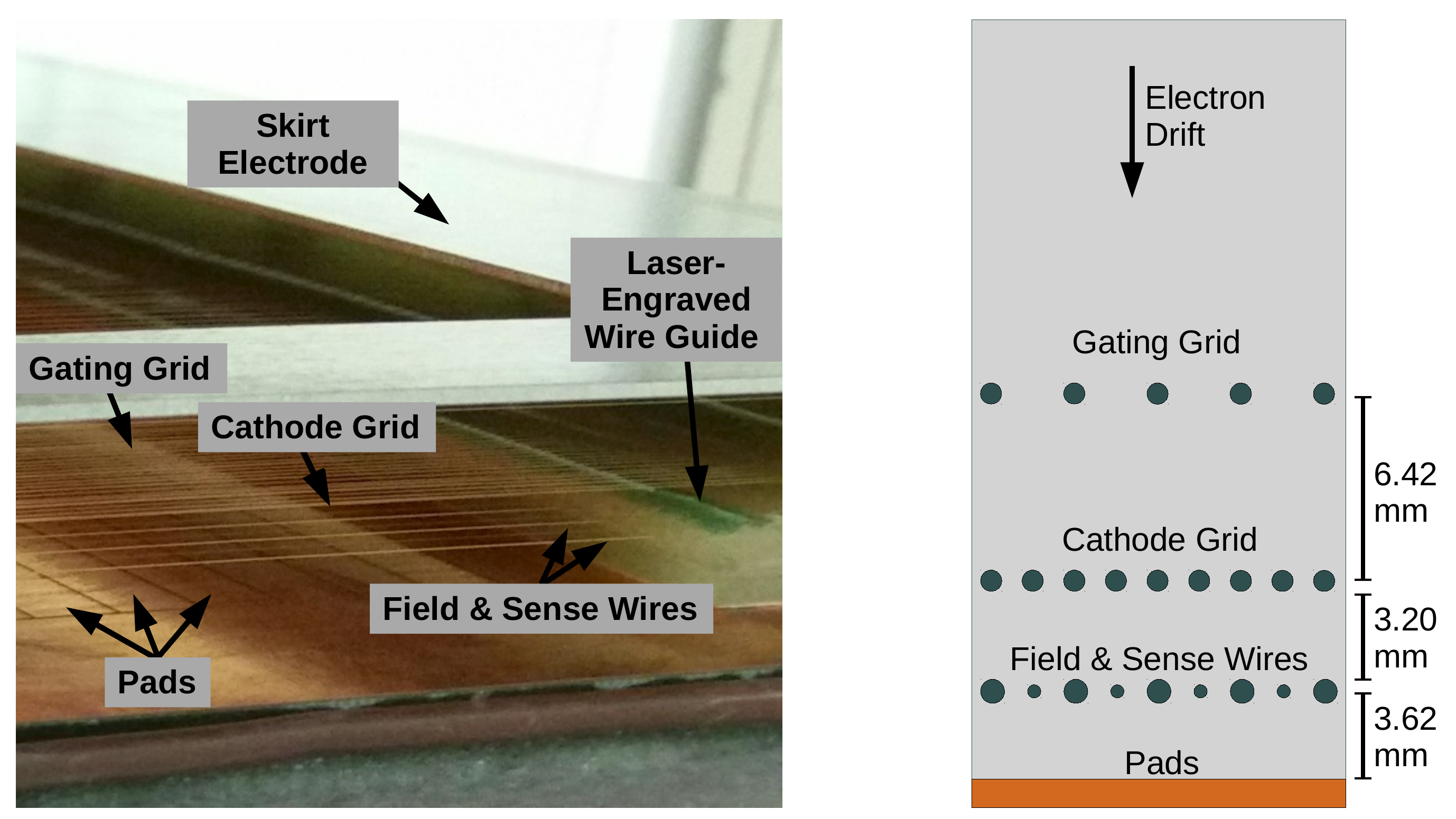}
  \end{center}
  \caption{(Color online) The structure of an FTPC amplification and 
    readout wire plane. Left: image of constructed MWPC and
    individual wireplanes. The Gating Grid, Cathode (Frisch) Grid, and the Field Wire / 
    Sense Wire Grid wire planes are stretched above the Pad Plane. 
    The precision of the Field Wire / Sense Wire pitch is set by a 
    laser engraved plastic wire guide. All sharp metal structures, 
    potentially sources of corona discharges, are covered by a 
    protective epoxy resin layer, as seen in the left figure.
    Right: Wireplane design cross-sectional view.
    }
  \label{figWirePlane}
\end{figure}

The primary geometric design of these planes was largely informed by the 
original NA49/NA61 TPC setting \cite{abgrall2014,afanasiev1999}, 
since the pad response function of the new FTPC system was required to 
conform to the standards of the existing upgraded NA61 TPC readout system 
\cite{laszlo2015}. The construction of the amplification and readout plane 
took place at the Innovative Detector Development Laboratory at the 
Wigner Research Centre for Physics, Budapest. 
A representative photograph of an FTPC wire plane with its annotated 
structure is seen in Figure~\ref{figWirePlane}.

Using a typical approach, from the drift direction it begins with a metal window 
(``skirt'') electrode set to transparent voltage, which defines the boundaries 
of the amplification and readout 
plane. Then, the wire plane of the gating grid follows: when open, it is set to a 
transparent voltage level. Otherwise when the chamber is idle, 
it is configured such that it blocks the drifting electrons from entering 
the amplification region and blocks ions from flowing into the active gas volume.
Below that, the Cathode (Frisch) grid follows, which defines 
the zero potential surface, closing the drift region. Finally, the electrons 
arrive at the sense wire/field wire amplification plane, where the actual 
electron multiplication happens. The field wires are kept at zero potential, which allows for
a lower amplification voltage on the sense wires.
Below that, a segmented pad plane is placed, 
reading out the mirror charges of the multiplied electrons around the sense wires. 

In order to relax the accuracy 
requirements on the wire winding process, the final wire geometry is fixed by 
a laser-engraved plastic wire guide with about $30\,\mu\mathrm{m}$ precision 
\cite{varga2013,varga2011}. The cathode flatness is ensured by gluing of 
the Pad Plane PCB onto the aluminium strongback with its face against a 
$20\,\mu\mathrm{m}$ precision optical table. 

In order to minimize the number of electronic readout channels, 
the FTPC2 and 3 chambers were subdivided into two sectors: for the downstream (upstream) half of FTPC2 (FTPC3), the pads are 67\% broader. This solution, 
while sparing readout channels, keeps the tracking lever 
arm and the number of $\mathrm{d}E/\mathrm{d}x$ sampling points intact. 
The main geometric parameters of the amplification and readout plane are 
listed in Table~\ref{tabWirePlane}. A typical operational sense wire voltage 
in Ar/CO${}_{2}$ (95:5) working gas was about $1.1\,\mathrm{kV}$.

\begin{table}[!h]
  \begin{center}
    \begin{tabular}{c|c|c|c|c|c|c}
      \emph{wire type}  & \emph{material} & \emph{thickness}  & \emph{manufacturer} & \emph{tension} & \emph{pitch}    & \emph{distance to} \\
                        &                 & [$\mu\mathrm{m}$] &                     & [$\mathrm{g}$] & [$\mathrm{mm}$] & \emph{Pad Plane} \,[$\mathrm{mm}$] \\
      \hline
      Sense Wire        & Au plated W     & $22$              & LUMA                & $19$           & $4$             & $3.62$ \\
      Field Wire        & Cu-Be alloy     & $114$             & CFW                 & $38$           & $4$             & $3.62$ \\
      Cathode Grid      & Cu-Be alloy     & $66$              & CFW                 & $29$           & $1$             & $6.82$ \\
      Gating Grid       & Cu-Be alloy     & $66$              & CFW                 & $29$           & $2$             & $13.24$ \\
    \end{tabular}
  \end{center}
  \begin{center}
    \begin{tabular}{c|c|c|c|c}
      \emph{chamber name}     & \emph{sensitive area}          & \emph{sensitive area}           & \emph{pad width}  & \emph{pad length} \\
                              & \emph{width} \,[$\mathrm{mm}$] & \emph{length} \,[$\mathrm{mm}$] & [$\mathrm{mm}$]   & [$\mathrm{mm}$] \\
      \hline
      FTPC1                   & $512$                          & $480$                           & $4$               & $40$ \\
      FTPC2 upstream sector   & $640$                          & $240$                           & $4$               & $40$ \\
      FTPC2 downstream sector & $640$                          & $240$                           & $6.6$             & $40$ \\
      FTPC3 upstream sector   & $640$                          & $240$                           & $6.6$             & $40$ \\
      FTPC3 downstream sector & $640$                          & $240$                           & $4$               & $40$ \\
    \end{tabular}
  \end{center}
  \caption{Main geometric and manufacturing parameters of the FTPC wire planes. 
           Top table: details of the wire grid geometry. 
           Bottom table: the geometric parameters of the readout pads.}
  \label{tabWirePlane}
\end{table}

\section{Performance}
\label{performance}

\begin{figure}[!h]
  \includegraphics[width=0.5\linewidth]{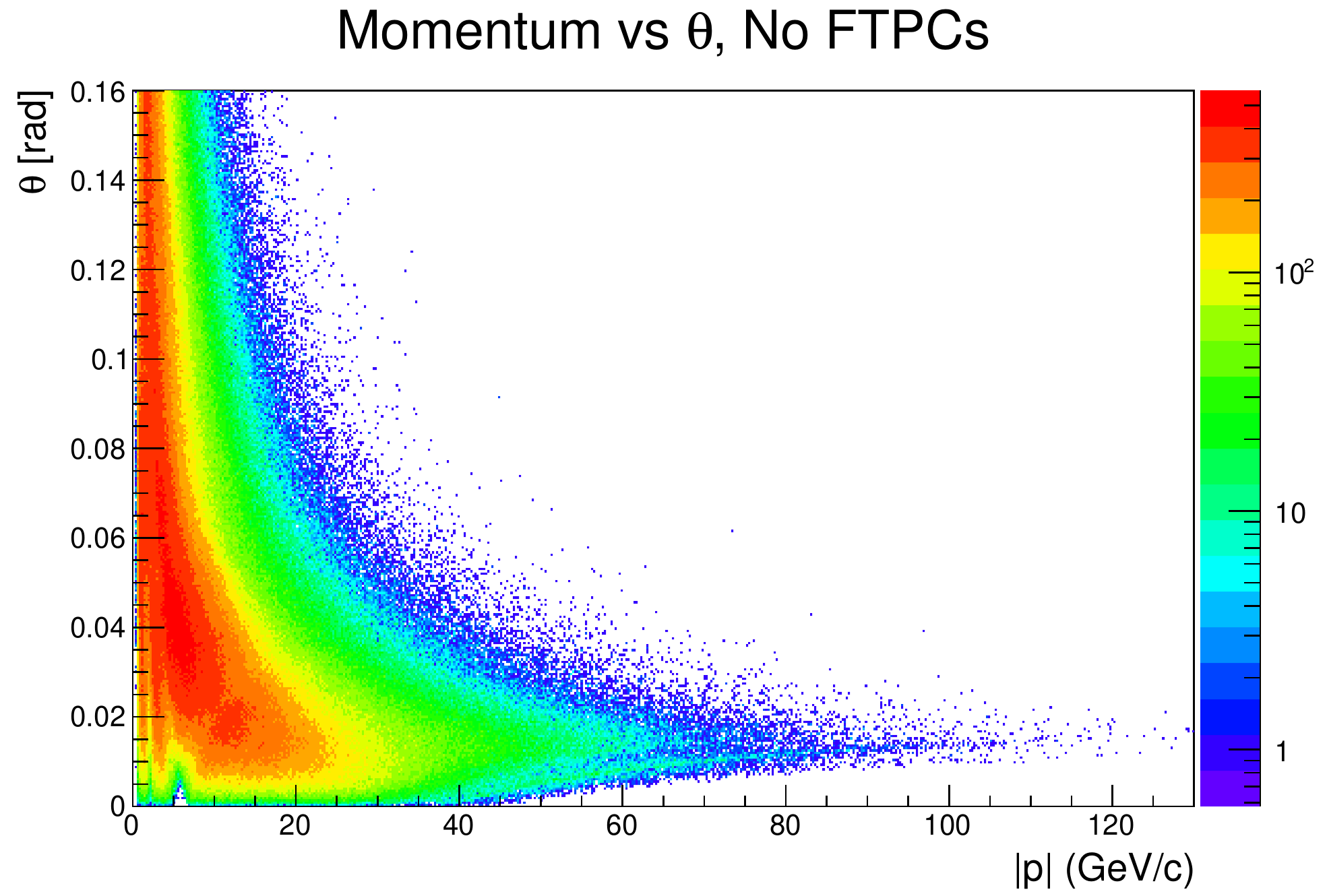}
  \includegraphics[width=0.5\linewidth]{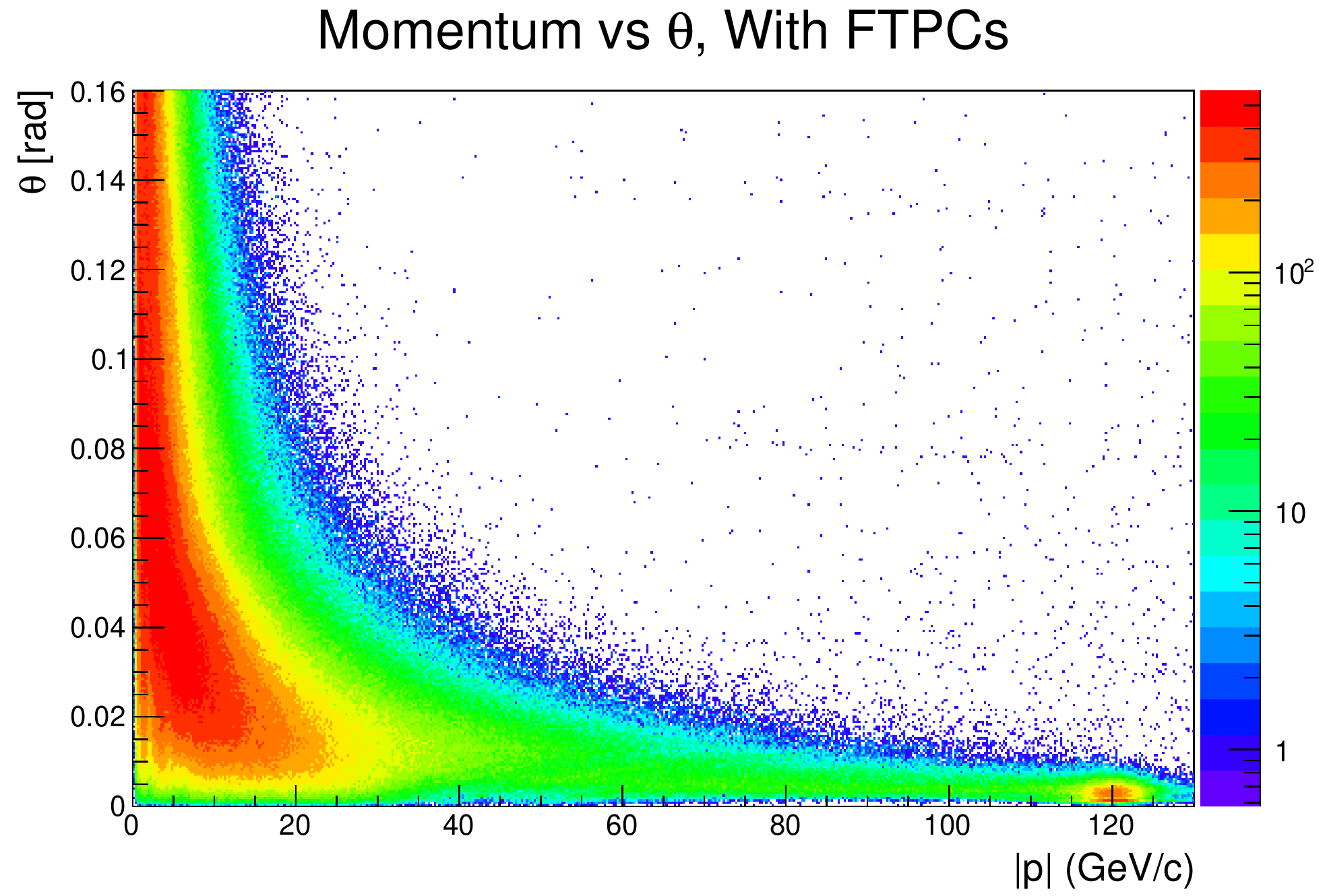}
  \caption{(Color online) Reconstructed track phase space coverage
    in $120\,\mathrm{GeV/c}$ proton-carbon interactions with and without
    inclusion of FTPC tracks. Left panel: tracks passing minimal
    cuts without inclusion of FTPC measurements. Right panel:
    tracks passing minimal cuts including FTPC tracks. Note the
    beam spot at $120\,\mathrm{GeV/c}$.
  }
  \label{figPhaseSpace}
\end{figure}

\begin{figure}[!h]
  \includegraphics[width=0.5\linewidth]{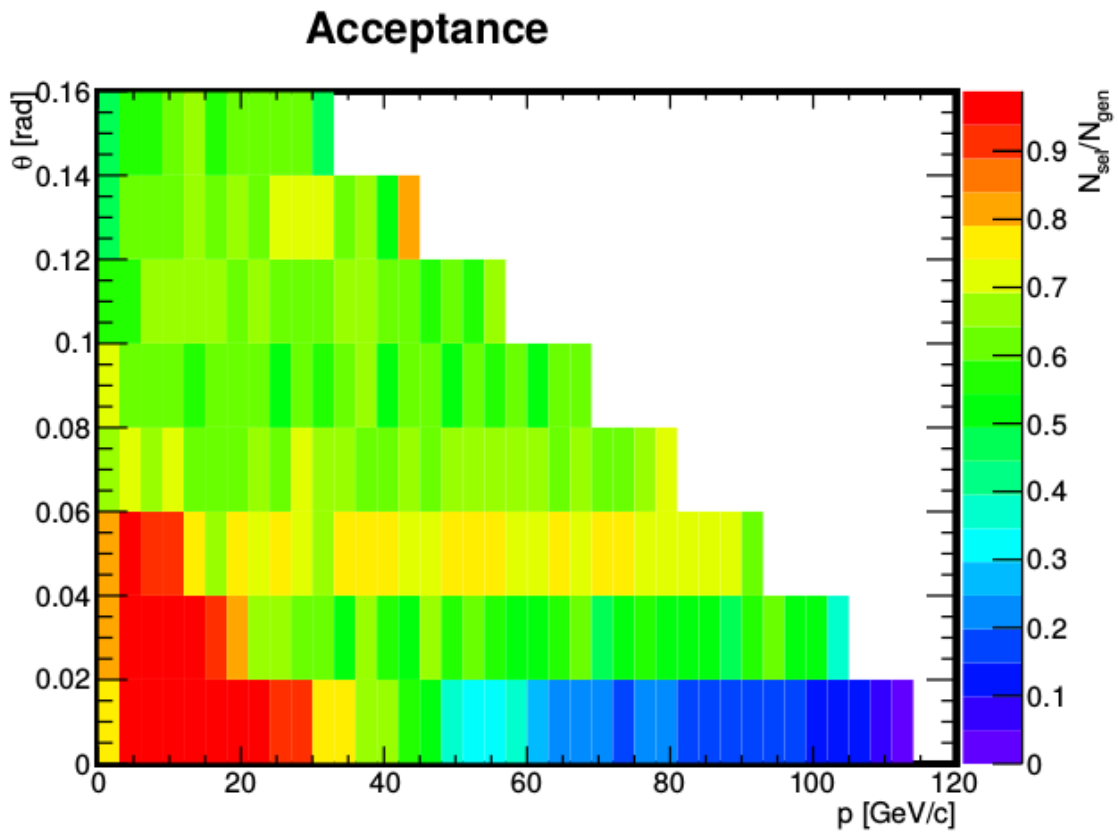}
  \includegraphics[width=0.5\linewidth]{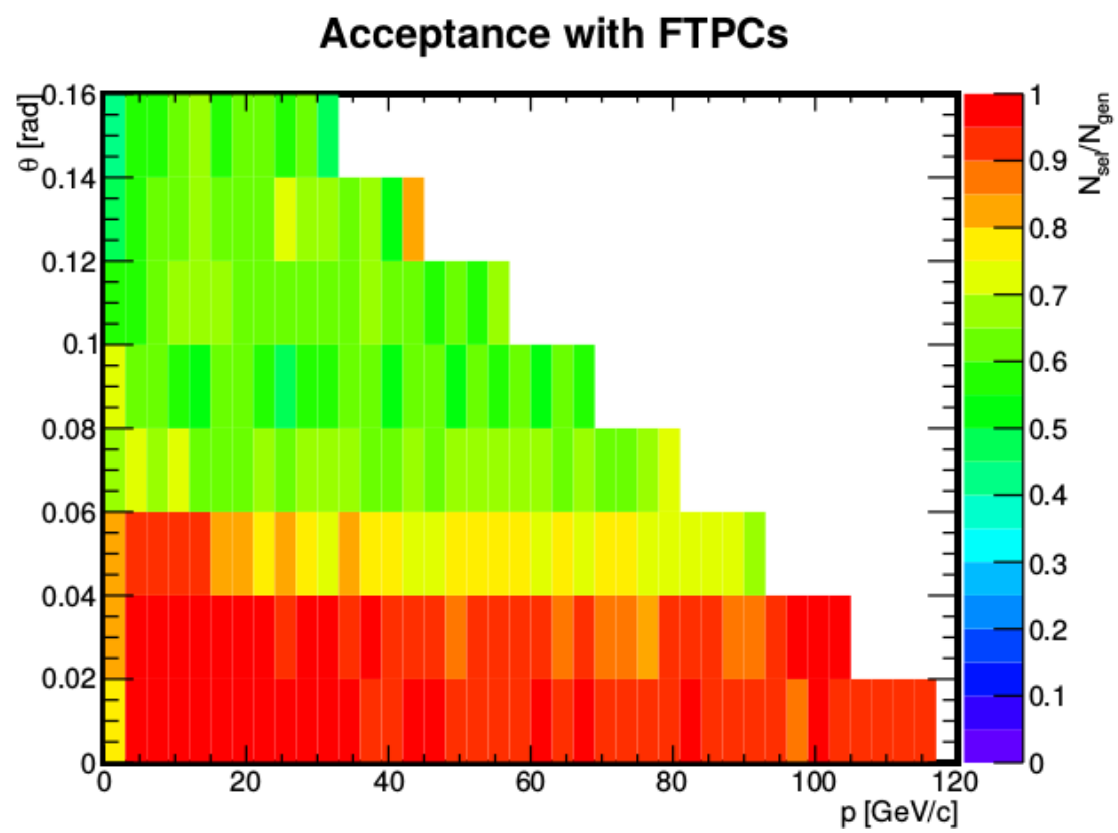}
  \caption{(Color online) Demonstration of NA61 detector acceptance
    in $120\,\mathrm{GeV/c}$ proton-carbon interactions before and after
    the addition of the FTPCs, as obtained from Monte Carlo simulation.
    $N_{\mathrm{gen}}$ denotes the number of generated tracks in a given
    $[p,\theta]$ bin, and $N_{\mathrm{sel}}$ means the number of tracks that
    generated sufficient measurement points to pass preliminary
    track selection in that bin. Left panel: NA61 acceptance before
    the integration of the FTPCs. Right panel: NA61 acceptance
    with FTPCs installed.
  }
  \label{figAcceptance}
\end{figure}

The addition of the FTPCs has resulted in several performance
increases in NA61. Primarily, the NA61 forward coverage has increased
dramatically, now extending up to the beam momentum. This can be
observed in Figure~\ref{figPhaseSpace}, showing the reconstructed
track phase space occupancy in proton-carbon interactions at
$120\,\mathrm{GeV/c}$ collected in 2017. The left hand panel shows
accepted tracks ignoring FTPC contributions, while the right hand
panel shows all accepted tracks including FTPC-GapTPC tracks. The
observed increase in coverage corroborates simulated acceptance plots
for identical beam and magnet configurations, shown in
Figure~\ref{figAcceptance}.

\begin{figure}[!h]
  \begin{center}
    \includegraphics[width=\linewidth]{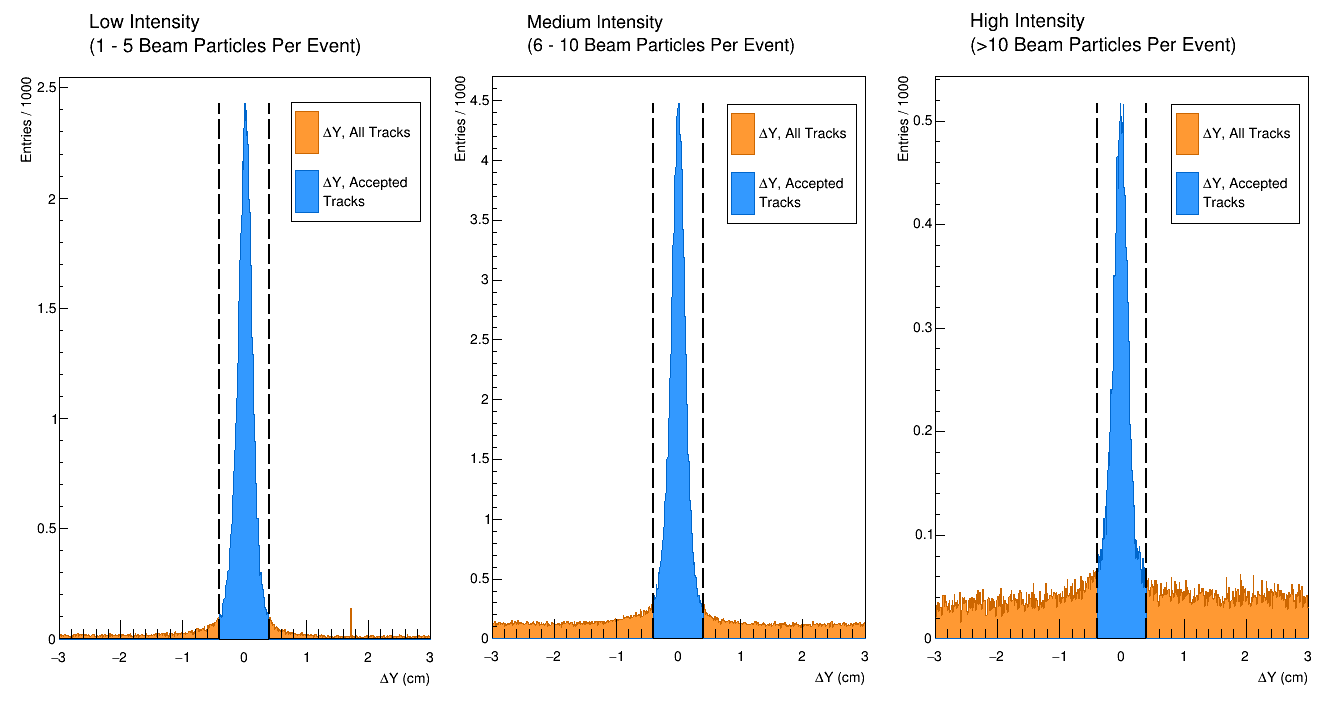}
  \end{center}
  \caption{(Color online) Background rejection with the tandem
    concept. The peak in each plot at $0\,\mathrm{cm}$ corresponds to
    in-time particles, while the continuous background (orange)
    corresponds to out-of-time particles. The out-of-time track
    background increases with beam intensity. The measured out-of-time
    fraction of tracks for each intensity (ratio of tracks in orange
    histogram to all reconstructed tracks) is 56\% (low-intensity),
    76\% (medium-intensity) and 83\% (high-intensity). For a standard
    TPC configuration, these tracks would be indistinguishable from
    the in-time tracks.}
  \label{figDY}
\end{figure}

As was foreseen in the design phase, the tandem-TPC concept
significantly reduces backgrounds related to out-of-time
tracks. Figure~\ref{figDY} demonstrates the background reduction
measured during the 2017 $120\,\mathrm{GeV/c}$ proton-carbon data
taking period, in which the typical beam intensity ranged from 1 to 25
beam particles per recorded event.  Figure~\ref{figDY} shows events
from this period separated into three intensity regimes: low beam
intensity (1-5 beam particles per event), medium beam intensity (6-10
beam particles per event), and high beam intensity (more than 10 beam
particles per event). Each intensity regime shows the distribution of
track position mismatch $\Delta{y}$ between the tandem pairs, where
the non-drift direction coordinate of the tracks are required to match
($\vert\Delta{x}\vert\leq 1\,\mathrm{cm}$). The peaks at $\Delta{y} =
0\,\mathrm{cm}$ correspond to in-time tracks, while the continuous
background corresponds to random matches of out-of-time track
pieces. As the beam intensity increases, so does the background of
out-of-time tracks.

This continuous background extends throughout the chamber in $y$,
i.e. along the drift coordinate, for the full $100\,\mathrm{cm}$ of
detector fiducial volume. If the tandem concept were not used, these
tracks would match both in $x$ and in $y$ and thus would provide a
significant background for data analysis. The method presented in
Figure~\ref{figDY} can also be used for purity estimation of the
out-of-time rejection of the tandem concept by extrapolating the
out-of-time background passing to the region under the $\Delta{y}$ cut
of the tandem pair.

Finally, we quantify the limitations of the tandem TPC system. Special cases may arise
when minimally-separated tracks in $y$ are joined to the same TPC track, or in the 
extreme case two ionizing particles may arrive too close in time for a TPC to 
resolve the two distinct tracks. Finding the threshold for this behavior corresponds to measuring the 
two-track time resolution of the TPC system. We perform this measurement using sequential
beam particles separated by just a few hundred nanoseconds. The timing 
reference of the beam particles were provided by the S1${}_{1}$ scintillator 
counter in the upstream beamline of NA61. A CAEN V1290N Multi-Hit
Time-to-Digital-Converter (MHTDC) records beam particle arrival time as 
measured on S1${}_{1}$, with a resolution of about $25\,\mathrm{psec}$ \cite{abgrall2014}, 
allowing for sub-nanosecond measurement of beam particle arrival time. The 
FTPC FEEs measure charge in $200\,\mathrm{nsec}$ time 
buckets in terms of drift time, and the clusterization algorithm for position 
estimation in the drift direction relies on at least three such samples. 
Thus, we expect the minimum two-track time resolution of the FTPCs to be of the order 
of $600\,\mathrm{nsec}$.

\begin{figure}[!h]
  \begin{center}
    \includegraphics[width=0.5\linewidth]{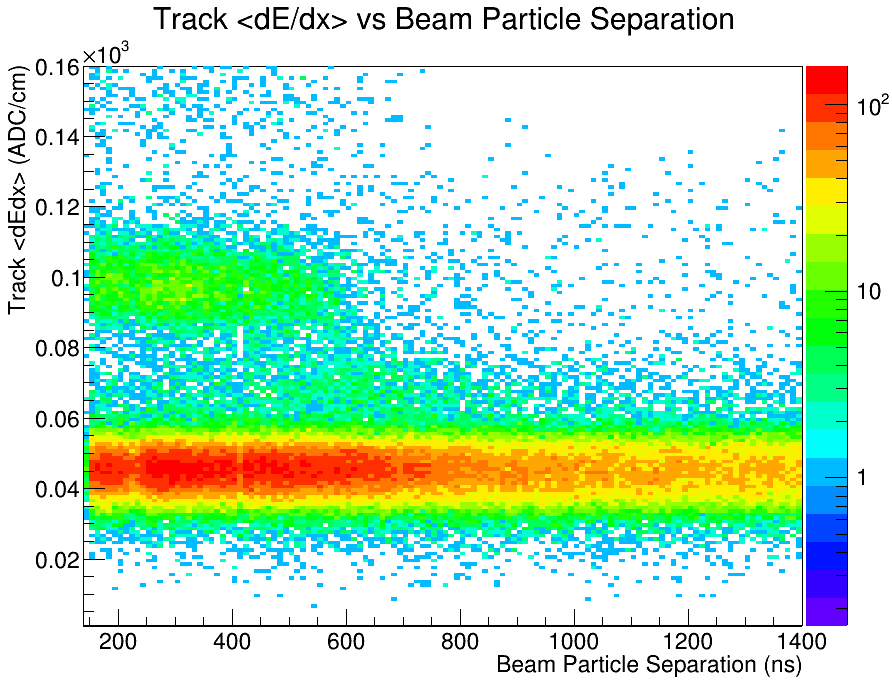}
  \end{center}
  \caption{(Color online) The limitation of the tandem-TPC concept: if two beam 
    particles arrive too close in time, the intrinsic two-track drift time resolution 
    of a TPC system is not enough for the separation of such out-of-time tracks, 
    even in a tandem-TPC configuration. In such situations, however, the typical 
    amplitude ($\mathrm{d}E/\mathrm{d}x$) response of the chamber indicates 
    anomalous amount of ionization signal: one observes a doubled signal amplitude 
    with respect to a typical MIP signal. Thus, such merged beam tracks can be 
    still labeled or rejected by their unusually large $\mathrm{d}E/\mathrm{d}x$.}
  \label{figtwotrack}
\end{figure}

In Figure~\ref{figtwotrack} we plot the beam particle
$\mathrm{d}E/\mathrm{d}x$ in the FTPCs vs the beam particle separation
time as seen by the MHTDC. It is seen that for larger separation 
times, a typical minimum ionization (MIP) $\mathrm{d}E/\mathrm{d}x$ signal 
is returned, corresponding to a single beam particle. Then, for separation times 
less then $600\,\mathrm{nsec}$ we observe a contribution corresponding to 
a $\mathrm{d}E/\mathrm{d}x$ signal of ${\sim}2\,$MIP, even for tracks 
required to satisfy tandem cuts. 
These ``double-$\mathrm{d}E/\mathrm{d}x$'' tracks are comprised of two beam 
particles closely spaced in time and merged together in their 
TPC response. 

This study shows that the limitation of the tandem-TPC based separation 
of out-of-time tracks comes mainly from the intrinsic two-track time resolution of a 
single TPC chamber, as expected. The ``double-$\mathrm{d}E/\mathrm{d}x$'' 
cut can, however, still be used to tag out-of-time contribution 
which are too close to be resolved by the tandem-TPC out-of-time rejection.

\section{Concluding remarks}
\label{conclusion}

In this paper the Forward TPC system of the NA61 experiment at CERN was 
described. The pertinent system was constructed in order to cover a 
previously uninstrumented part of the NA61 phase space in the very forward 
region, i.e.\ in and around the beam line. Since the FTPC system 
is installed in the beam region, recognition of background tracks originating 
from previous events (out-of-time particle tracks) is essential. In order to 
achieve this, we employed a novel technique, using alternating drift field 
directions, which we called a tandem-TPC concept. In such a setting, 
the tracks of out-of-time particles already start to drift in opposite 
directions in the subsequent chambers, thus the tracks of these will 
be discontinuous throughout the system. 

As a primary objective, we have demonstrated that this new 
tandem-TPC concept excels at rejecting out-of-time background tracks, 
in a relatively high-intensity beam (${\sim}100\,\mathrm{kHz}$). 
At the same time, due to the nature of the TPC concept, a good tracking and 
$\mathrm{d}E/\mathrm{d}x$ capability, as well as low material budget 
and cost effectiveness were also achieved. The tandem-TPC
concept could also perform well at other relatively high-intensity beam 
facilities, and could satisfy the above requirements.

\section*{Acknowledgements}

We would like to thank to Leszek Ropelewski and Eraldo Oliveri of CERN for kindly 
providing laboratory space and basic facilities for the joining procedure 
of the FTPC field cage and the wire plane at CERN. 
We thank to the support and help from all the members of the CERN 
NA61/SHINE Collaboration.
This work was supported by the Hungarian Scientific Research Fund 
(NKFIH 123842-123959), the Momentum (``Lend\"ulet'') program of the 
Hungarian Academy of Sciences, the Hungarian National Research, Development 
and Innovation Office (NKFIH T\'ET\_16\_CN-1-2016-0008), and the U.\ S.\ Department
of Energy (grant DE-SC0010005).

\end{document}